\documentclass{iopart}             
\usepackage[T1]{fontenc}
\usepackage[applemac]{inputenx}
\usepackage{iopams}
\usepackage{graphicx}
\usepackage{url,graphics,epsfig}
\usepackage{amssymb}
\usepackage[breaklinks=true,colorlinks=true,linkcolor=blue,urlcolor=blue,citecolor=blue]{hyperref}

\usepackage{tikz}
\usepackage{cite}

\begin{document}
\jl{2}
%
%
%
\def\etal{{\it et al~}}
\def\newblock{\hskip .11em plus .33em minus .07em}
%
%
%
%
%
\setlength{\arraycolsep}{2.5pt}             

\title[Photoionization of  W$^{+}$ ions] {Single-photon single ionization of W$^{+}$ ions:  experiment  and theory}

\author{A M\"{u}ller$^1\footnote[1]{Corresponding author, E-mail: Alfred.Mueller@iamp.physik.uni-giessen.de}$,
		S Schippers$^{1,2}$, J Hellhund$^{1,4}$, K Holste$^2$,\\  A L D Kilcoyne$^3$, R A Phaneuf$^4$,\\
		C P Ballance$^{5,6}$, and B M McLaughlin$^{6,7}\footnote[2]{Corresponding author, E-mail: b.mclaughlin@qub.ac.uk}$}

\address{$^1$Institut f\"{u}r Atom- ~und Molek\"{u}lphysik,
                         Justus-Liebig-Universit\"{a}t Giessen, 35392 Giessen, Germany}

\address{$^2$I. Physikalisches Institut,
                           Justus-Liebig-Universit\"{a}t Giessen, 35392 Giessen, Germany}

\address{$^3$Advanced Light Source, Lawrence Berkeley National Laboratory,
                          Berkeley, California 94720, USA }

\address{$^4$Department of Physics, University of Nevada,
                          Reno, NV 89557, USA}

\address{$^5$Department of Physics,  206 Allison Laboratory,
                            Auburn University, Auburn, AL 36849, USA}

\address{$^6$Centre for Theoretical Atomic, Molecular and Optical Physics (CTAMOP),
                          School of Mathematics and Physics, The David Bates Building, 7 College Park,
                          Queen's University Belfast, Belfast BT7 1NN, UK}

\address{$^7$Institute for Theoretical Atomic and Molecular Physics,
                          Harvard Smithsonian Center for Astrophysics, MS-14,
                          Cambridge, MA 02138, USA}
%
%
%

\begin{abstract}
Experimental and theoretical results are reported  for photoionization of  Ta-like (W$^{+}$)  tungsten ions. Absolute cross sections were measured in the energy range 16 to 245~eV employing the photon-ion merged-beam setup at the Advanced Light Source in Berkeley. Detailed photon-energy scans at 100 meV bandwidth were performed in the 16 to 108 eV range. In addition, the cross section was scanned at 50~meV resolution in regions where fine resonance structures could be observed.  Theoretical results were  obtained from a Dirac-Coulomb R-matrix approach. Photoionization  cross section calculations were performed for singly ionized atomic  tungsten ions in their $5s^2 5p^6 5d^4({^5}D)6s \; {^6}{\rm D}_{J}$, $J$=1/2, ground level and the associated excited metastable levels with $J$=3/2, 5/2, 7/2 and 9/2.  Since the ion beams used in the experiments must be expected to contain long-lived excited states also from excited configurations, additional cross-section calculations were performed for the second-lowest term, $5d^5 \; ^6{\rm S}_{J}$, $J$=5/2, and for the $^4$F  term, $5d^3 6s^2 \;  ^4{\rm F}_{J}$, with  $J$ = 3/2, 5/2, 7/2 and 9/2. Given the complexity of the electronic structure of W$^+$ the calculations reproduce the main features of the experimental cross section quite well.
\end{abstract}

%
%

\pacs{32.80.Fb, 31.15.Ar, 32.80.Hd, and 32.70.-n}

\vspace{0.5cm}
\begin{flushleft}
Short title: Valence shell photoionization of singly ionized atomic tungsten ions\\
\vspace{1.0cm}
\end{flushleft}
%
\maketitle
%
%
%
\section{Introduction}
Tungsten presently receives substantial scientific interest because of its importance in nuclear-fusion research. Due to its high thermal conductivity, its high melting point, and its resistance to sputtering and erosion tungsten is the favoured material for the wall regions of highest particle and heat load in a fusion reactor vessel~\cite{Neu2013}. Inevitably, tungsten atoms and ions are released from the walls and enter the plasma. With their high atomic number, $Z=74$, they do not become fully stripped of electrons and therefore radiate copiously, so that the tolerable fraction of tungsten impurity in the plasma is at most 2$\times$10$^{-5}$~\cite{Neu2003}. Understanding and  controlling tungsten in a plasma requires detailed knowledge about its collisional and spectroscopic properties. Although not directly relevant to fusion,  photoionization of tungsten atoms and ions is interesting because it can provide details about spectroscopic aspects and, as time-reversed photorecombination, helps to better understand one of the most important atomic collision processes in a fusion plasma, electron-ion recombination.

R-matrix theory is a tool to obtain information about electron-ion and photon-ion interactions in general. Electron-impact ionization and recombination of tungsten ions have been  studied experimentally~\cite{Mueller2015b,Rausch2011a,Stenke1995b,Stenke1995c,Schippers2011b,Krantz2014,Spruck2014}
while there are no detailed measurements on electron-impact excitation of tungsten atoms in any charge state. Thus, the present study on photoionization of these complex ions and the comparison of experimental data with R-matrix calculations provides benchmarks and guidance for future theoretical work on electron-impact excitation.

Photoabsorption by neutral tungsten atoms in the gas phase has been studied experimentally by Costello et al. employing the dual-laser-plasma technique~\cite{Costello1991}. A few years later the production of W$^+$ and W$^{2+}$ photo-ions from tungsten vapor was observed by Sladeczek et al.~\cite{Sladeczek1995}. However, no experimental data have been available in the literature for tungsten ions prior to the present project. Direct photoionization of W$^{q+}$ ions is included in the calculations by Trzhaskovskaya et al. \cite{Trzhaskovskaya2010} as time-reversed radiative recombination but is expected to be only a small contribution to the total photoionization cross section. Theoretical work using  many-body perturbation theory (MBPT) has been carried out by Boyle et al.~\cite{Boyle1993} for photoionization of neutral tungsten atoms. Theoretical treatment of photoabsorption using a relativistic Hartree-Fock (RHF) approach was reported by Sladeczek et al.~\cite{Sladeczek1995} in conjunction with their experiments. Very recently,  Ballance and  McLaughlin have  carried out large-scale R-matrix calculations for the neutral tungsten atom~\cite{Ballance2015a} using the Dirac-Coulomb R-matrix approximation implemented in the DARC codes. The present study is the first investigation on photoionization of tungsten ions and addresses singly charged W$^+$. Preliminary reports on our ongoing tungsten photoionization project were presented at conferences~\cite{Mueller2011a,Mueller2012,Mueller2014c} previously.

The ground level of the Ta-like W$^{+}$ ion is $5p^65d^4(^5D)6s~^6{\rm D}_{1/2}$ with an ionization potential of (16.37 $\pm$ 0.15)~eV~\cite{NIST2014}. One must assume that along with the $^6{\rm D}_{1/2}$ ground level, the excited ground-configuration fine-structure levels $^6{\rm D}_{J}$ with $J$=3/2, 5/2, 7/2 and 9/2 at excitation energies below 0.8~eV~\cite{NIST2014}, respectively, are also populated in an ion source that produces W$^{+}$ by electron-impact ionization of neutral tungsten. Also the lowest levels of the first excited $5d^5$ and $5d^36s^2$ configurations have excitation energies below 2~eV and are likely populated in the ion-source plasma. The energetically lowest configurations $5p^6 5d^4 6s$, $5d^5$ and $5d^36s^2$ all have even parity and, hence, all the 118 excited levels within these configurations are long-lived because electric dipole transitions between any of these levels are forbidden. Any strong signal in the experimental photoionization spectrum below the threshold of (16.37 $\pm$ 0.15)~eV would indicate the presence of metastable excited states in the parent ion beam, most likely within the $5p^65d^4(^5D)6s~^6{\rm D}$, $5p^65d^5~^6{\rm S}$ and $5p^65d^3 6s^2~^4{\rm F}$ terms.

The direct and resonant photoionization  processes  occurring in the present energy range up to 245~eV for the interaction of a single
photon with  the ground-state and the lowest metastable configurations of the Ta-like tungsten ion comprise removal or excitation of
either a $4f$, $5s$, $5p$, $5d$ or a $6s$ electron. For the theoretical description of W$^+$ photoionization suitable
target wave functions have to be constructed that allow for promotions of electrons from these subshells to all contributing excited states.
This is challenging for a low-charge ion such as W$^+$ but becomes simpler for the ions in higher charge states due to the increased
effect of the Coulomb charge of the target and the slight reduction in the R-matrix box size.

This paper is structured as follows. Section 2 details the experimental procedure used.
Section 3 presents a brief outline of the theoretical work. Section 4 presents a discussion of the
results obtained from both the experimental and theoretical methods.
Finally in section 5 conclusions are drawn from the present investigation.
An appendix has been added to describe corrections of measured photoionization
cross sections for effects of higher-order radiation present in the photon beam particularly at low photon energies.
%
%
%
%
%

\section{Experiment}\label{sec:exp}

The measurements on photoionization of W$^{+}$ ions were carried out at the Ion-Photon Beam (IPB) endstation of beamline 10.0.1.2 at the Advanced Light Source ALS in Berkeley, California, USA. The general layout of the experimental setup and the procedures  employed have been described previously by Covington \etal~\cite{Covington2002a}. Technological developments since these early  measurements have been discussed recently by M\"{u}ller \etal~\cite{Mueller2014b}. An overview of the experiment is presented here and aspects specific to the present measurements are discussed in detail.

Ta-like W$^+$ ions were produced from W(CO)$_6$ vapour in an electron-cyclotron-resonance (ECR) ion source. The ions were extracted and accelerated to ground potential by a voltage of 6~kV forming a beam of ions that was composed of a complex mixture of electrically charged fragments of the initial tungsten hexacarbonyl molecules. The ion beam was steered and focused electrostatically to the entrance aperture of a 60$^\circ$-bending-angle dipole magnet which separated the ions with respect to their momentum per charge. A beam of isotopically pure $^{186}$W$^+$ ions was selected by appropriately choosing the field of the analyzer magnet. The $^{186}$W$^+$ ion beam emerging from the exit aperture of the magnet was deflected by a hemispherical electrostatic 90$^\circ$ deflector, the ''merger'',
onto the axis of the counter-propagating narrow-bandwidth photon beam provided at the beamline.

W$^{2+}$ product ions formed as a result of interactions of single photons with single W$^{+}$ parent ions were separated from the parent ion beam by a second magnet positioned downstream of the interaction region (upstream of the photon beam). This so called ''demerger'' magnet deflected the product ion beam by  45$^\circ$ so that it could enter a single-particle detector unit while the parent ion beam, deflected by half that angle, was collected by a large Faraday cup. The detector unit consists of a hemispherical electrostatic 90$^\circ$ deflector, a variable-size aperture based on a four-jaw movable-slit system and a single-particle detector with almost 100\% efficiency~\cite{Fricke1980a,Rinn1982}. The additional deflection served to decouple the detector from stray particles and photons produced in the deflection plane of the demerger magnet as well as to sweep the product beam across the detector in a direction orthogonal to that by the demerging magnet to ensure  complete detection of products.

For optimizing W$^{2+}$ signal count rates the ion beam was tuned for maximum overlap with the photon beam positioned on the axis of the merging section of the apparatus between the ''merger'' deflector and the ''demerger'' magnet. The overlap of the ion and photon beams was quantified by using three slit scanners at the entrance, exit, and middle position of the photon-ion interaction region which was defined by a metal drift tube with entrance and exit apertures.  Each slit scanner could probe the overlapping beams in two directions perpendicular to one another thus providing access to the form factor $F(z)$~\cite{Phaneuf1999,Covington2002a} at each position z. The total form factor $\mathfrak{F} = \int{F(z)dz}$ characterizing the three-dimensional overlap of the two beams in the interaction region was then obtained by interpolation of the position-dependent form factors $F(z)$ considering that the trajectories of ions and photons within the interaction region are straight lines.

For maximum reduction of background arising from collisions of W$^+$ parent ions with residual-gas particles along the whole merging section the pressure in this section was kept at a minimum. Operating pressures ranged from about 3 to 5$\times 10^{-10}$~mbar. For separation of the signal of photoionized tungsten ions from background of collisionally-produced W$^{2+}$ ions the photon beam was chopped at approximately 6~Hz during the data acquisition  so that the background and signal-plus-background count rates could be measured separately. By subtracting the former from the latter the instantaneous signal count rate was recovered. Total signal counting times up to hundreds of seconds were used to obtain suitable levels of statistics.

Two modes of operation were applied to obtain information about the photoionization cross section of W$^+$ ions. In ''spectroscopy mode'' the voltage on the interaction region was set to 0~V so that signal could be collected from the whole merging section (approximately 1.4~m length) between the ''merger'' and ''demerger'' units. At constant but unknown beam overlap the energy dependent W$^{2+}$ signal count rate $R(E_\gamma)$ (obtained by chopping the photon beam) was recorded as a function of photon energy $E_\gamma$. These energy-scan measurements were carried out at 100~meV constant resolution covering an energy range from 16 to 108~eV in 50~meV steps (at energies below 27.1~eV the beamline provides beams with lower bandwidths; at a fixed maximum exit slit width of 1300~$\mu$m the beamline resolution gradually drops  from 100~meV at 27.1~eV to about 36~meV at 16~eV.) At the same time the primary-ion and photon fluxes, $\dot{N}_i$ and $\dot{N}_{\gamma}$, respectively, were recorded so that a normalized relative cross section $R/(\dot{N}_i \dot{N}_{\gamma})$ could be obtained for each energy step. As already mentioned, the ion beam was collected in a large Faraday cup and its electrical current $I_i = e \dot{N}_i$  was measured with a precision electrometer with $e$ being the elementary charge. The photons were collected on the sensitive area of a calibrated photodiode. The photocurrent $I_{\gamma} = e Q(E_\gamma) \dot{N}_{\gamma}$ was measured with a similar electrometer where $Q(E_\gamma)$ is the known conversion efficiency of the photodiode, i.e., the number of electrons per incident photon. The relative cross section function obtained by energy-scan measurements was normalized afterwards to absolute cross sections measured in a separate phase of the experiment.

In ''absolute mode'' absolute cross-sections $\sigma(E_\gamma)$  at selected photon energies $E_\gamma$ were determined from
\begin{equation}
\label{Eq:xsec}
\sigma(E_\gamma) = \frac{R(E_\gamma) q e^2 v_i Q(E_\gamma)}{I_i I_{\gamma} \eta \mathfrak{F}(E_\gamma)}
\end{equation}
where  $q$ is the charge state of the primary ions, $v_i$ the primary ion velocity and $\eta$ the detection efficiency of the product-ion detector system. The quantities $R$, $I_i$ and $I_{\gamma}$ are readily obtained from the experiment, $Q(E_\gamma)$ and $\eta$ are known from separate calibration measurements. The determination of the conversion efficiency $Q(E_\gamma)$ was  performed with reference to a photodiode calibrated by
NIST~(for this service see url http://www.nist.gov/pml/div685/calibrations.cfm).  In order to determine $\mathfrak{F}(E_\gamma)$  the interaction length has to be experimentally defined. For this purpose the photoionized (W$^{2+}$) ions produced in the interaction region were energy-tagged by applying a voltage of +500~V to the metal tube defining the interaction region. Parent W$^{+}$ ions were decelerated to an energy of 5.5~keV when entering the interaction region. Product W$^{2+}$ ions generated inside the drift tube were accelerated to 6.5~keV when leaving the interaction region thus gaining a net total energy of 6.5~keV compared to the 6~keV of those W$^{2+}$ ions which were formed outside of the interaction region. The interaction length could thus be defined as the ''inside'' length of the potential barrier applied to the drift tube which was (29.4$\pm$0.6)~cm. The downstream demerger and detection unit can easily separate W$^{2+}$ ions with energies of 6 and 6.5~keV energy. The energy-tagged 6.5~keV W$^{2+}$ ions unambiguously arise from the defined interaction length of 29.4~cm and are associated with the measured total form factor $\mathfrak{F}$.

Typical ion currents in collimated beams used for the absolute measurements were 15~nA. Photon fluxes strongly depended on the energy $E_\gamma$ and the desired energy bandwidth. For consistency, most absolute measurements were performed at 45~meV bandwidth defined by suitable settings of the monochromator slits. In a few cases  bandwiths of 50~meV and 70~meV were employed which did not have a significant influence on the cross section results. This was expected  given the fact that the measurements were carried out at energies where there were no obvious resonances in the spectrum. Numbers for the photon flux were about $1\times 10^{11}$~s$^{-1}$ at 16~eV,
$3\times 10^{14}$~s$^{-1}$ at 40~eV, $1.4\times 10^{13}$~s$^{-1}$ at 120~eV, and $7\times 10^{12}$~s$^{-1}$ at 245~eV. Total beam overlaps $\mathfrak{F}$ varied in the absolute measurements between about  200~cm$^{-1}$ and 500~cm$^{-1}$. Signal count rates varied from 40~kHz at 40 eV to 0.8~Hz at 245~eV. Background count rates were near 1~Hz.

The present measurements were extended to energies down to 16~eV because the ground-level ionization energy is expected to be at (16.37 $\pm$ 0.15)~eV~\cite{NIST2014}. This energy is stretching the range of applicability of the available low-energy-grating monochromator of the beamline. A serious problem at such low energies is the rapidly decreasing photon flux while the fraction of higher-order radiation is increasing  relative to the first-order flux.
While corrections for higher-order effects on cross sections measured at the IPB endstation of BL 10.0.1.2 at the ALS have been applied previously~\cite{Lu2006a,Esteves2011}, a more extensive assessment of such effects was in order  due to the increased relative magnitudes of these contributions. Their effect is further increased in the case of W$^+$ because the photoionization cross sections at twice and three times the ground-state ionization threshold energy are relatively large. Fractions of higher-order radiation were determined by separate experiments which are described in the Appendix. From the results of that investigation corrections were derived for the measured apparent cross sections. The corrections are large for the lowest photon energies reaching almost a factor of 2 at 16~eV. At 20~eV the correction is down to 37\% and at 25~eV  amounts to only 17\% which is less than the typical uncertainty of absolute measurements of photoionization cross sections at the ALS IPB endstation.

The error budget of cross section measurements has recently been discussed by M\"{u}ller \etal~\cite{Mueller2014b} who estimated a systematic uncertainty of 19\% for cross sections determined by the use of Eq.~\ref{Eq:xsec}. This uncertainty includes the uncertainty of the photodiode calibration curve. It is worth mentioning that the quoted possible error of the calibration is only about 1\% in the energy range 37.5~eV to 170~eV. This precludes significant changes of the photoionization spectrum due to uncertainties of the calibration curve - at least in the energy range where this uncertainty is negligibly small.

The energies investigated in the publication by M\"{u}ller \etal~\cite{Mueller2014b} ($E_\gamma \geq 95$~eV)  were far beyond the region where significant corrections for higher-order effects would be required. In the present study, however, the uncertainty of the higher-order corrections has to be considered. Moreover, the photodiode used in the present experiment had already been exposed to substantial photon doses. Therefore the photon flux measured with the present photodiode was compared with the response of a calibrated pristine photodiode covering the whole energy range investigated here for which three different gratings were used in the monochromator. The comparison showed an average reduction of the diode conversion efficiency of 5\% for the heavily used diode. This fatigue effect was corrected for. On the basis of the observed fluctuations the additional relative uncertainty of the photon flux measurements with the two photodiodes was estimated to be 9\% and the associated absolute uncertainties were  included in quadrature in the total systematic uncertainty. In addition, an uncertainty of 50\% of the difference between the uncorrected and the higher-order-corrected cross section (see Appendix) was assumed. The associated absolute uncertainty as well as the statistical uncertainty of the measured count rate $R$ was added in quadrature  in order to obtain an estimate of the total possible error of the cross section. Total absolute error bars are shown for absolute cross sections in Sec.~\ref{Sec:Results}.

In addition, the photon-energy calibration of the present measurements has uncertainties. The counterpropagation of the photon and ion beams results in small Doppler effects of $2.5\times10^{-4}$ of the actual photon energy. Corrections of $E_\gamma$ are thus at most 62~meV at 245~eV. The uncertainties of such corrections are negligible. The energy axis was cross-calibrated to known resonance positions in the photoionization of Ar$^+$ ions. Since there are no sharp cross section features in the investigated energy range the calibration of the energy axis was not a prime issue. We estimate a maximum uncertainty of the energy axis of at most $\pm$100~meV in the energy range 16 to 80~eV where the only structures in the cross section occur.

As mentioned in the introduction, the 118 excited levels (plus the ground level) in the lowest-energy configurations $5p^6 5d^4 6s$, $5d^5$ and $5d^36s^2$ of W$^{+}$ are all  expected to be long lived. In principle, they can all be populated by energetic electron collisions in the plasma of the ECR ion source~\cite{Mueller2015b}. It is not \textit{a priori} evident, however, which levels are populated at what relative weight. Given the small energy splitting of fine-structure levels within a given term, it is reasonable to assume that all levels within that term are populated and that the population is statistical, i.e., it follows the statistical weight of each level within the term. In this context it may be worth mentioning that measurements on the photoionization of W$^{+}$ ions have been carried out over a time range of 5 years during 4 different beamtimes at the ALS. Energy-scan spectra in certain energy ranges were taken repeatedly and with different operation modes of the ion source. Nevertheless, the scan measurements and particularly the cross section features at certain energies were always reproducible. This is a strong indication for consistent and reproducible sets of initial-level populations in the primary ion beams used in all those experiments. The photoionization cross section results suggest that there was a strong, maybe even dominant, contribution from ions in the lowest-energy term, the $^6$D term,  within the $5p^65d^46s$ ground-state configuration.
%
%
%
%
%
\section{Theory}\label{sec:Theory}

For comparison with the measurements  made at the ALS, state-of-the-art theoretical methods using highly correlated wavefunctions  were applied that include relativistic effects. An efficient parallel version \cite{Ballance2006}  of the DARC \cite{Norrington1987,Wijesundera1991,darc} suite  of codes was applied which has been developed \cite{Fivet2012,McLaughlin2012a,McLaughlin2012b} to address electron and photon interactions with atomic systems providing for hundreds of levels and thousands of scattering channels. These  codes are presently running on a variety of parallel high performance computing architectures world wide \cite{McLaughlin2014a,McLaughlin2014b}. Recently, DARC calculations on photoionization of trans-Fe elements were carried out for Se$^{+}$, Kr$^{+}$, Xe$^{+}$, and Xe$^{7+}$ ions \cite{McLaughlin2012a,McLaughlin2012b,Hinojosa2012,Mueller2014b} showing suitable agreement with high resolution ALS measurements.

Photoionization cross section calculations on Ta-like W$^+$ ions were performed for 10 selected levels in the ground and the lowest excited configurations $5s^25p^65d^46s$, $5s^25p^65d^5$  and $5s^25p^65d^36s^2$. The atomic structure calculations for the W$^{2+}$ product ion were carried out using the GRASP code \cite{Dyall1989,Parpia2006,Grant2007}. We included 573 levels in our close-coupling calculations resulting from the 6 configurations $5s^25p^65d^4$,  $5s^25p^65d^36s$,  $5s^25p^65d^36p$,  $5s^25p^65d^36d$, $5s^25p^65d^26s^2$ and $5s^25p^55d^5$ to represent  the atomic structure of the W$^{2+}$ residual ion. The 573-level approximation is the simplest approximation which allows for the opening of both the $5d$ and $5p$ subshell. Alternatively, a 449-level model based on the 6 configurations $5s^25p^65d^4$,  $5s^25p^65d^36s$,  $5s^25p^65d^36p$,  $5s^25p^65d^36d$, $5s^25p^65d^26s^2$ and $5s^25p^65d^26s6p$ was used to see possible differences in the representation of the atomic structure of W$^{2+}$ levels.

In Table 1 we show a comparison of the 449-level and 573-level target approximations obtained from the GRASP code with the tabulated values from NIST. Note the difficulty of accurately describing the energy levels in the two approximations for near neutral states of tungsten. Larger target expansions (allowing for two-electron promotions to higher lying residual orbitals, and the opening of further inner shells) would naturally bring the theoretical results in better agreement with experiment but would be prohibitive for scattering and photoionization calculations. In other words, extending the basis set for describing photoionization of W$^+$ with its open $5d$ and $6s$ subshells by including more configurations of the W$^{2+}$ product ion would quickly increase the number of levels and thus require a computational effort that would go well beyond the limitations set by presently available computing resources. Therefore, the 573-level approximation has to be considered a reasonable compromise between adequate representation of the tungsten ion structure and feasibility of the photoionization computations at the present technical limit of {\em ab initio} close-coupling treatment.

The cross section calculations for this 573-level model were carried out in  the Dirac-Coulomb approximation using the DARC codes \cite{McLaughlin2012a,McLaughlin2012b} for photon energies from the ionization thresholds up to 150~eV. The R-matrix boundary radius of 10.88 Bohr radii  was sufficient to envelop the radial extent of all the n=6 atomic orbitals of the residual W$^{2+}$ ion. A basis of 16 continuum orbitals was sufficient to span the  photon energy range chosen for the calculations. Since dipole selection rules apply,  total ground-state photoionization cross sections require only  the  bound-free dipole matrices, $2J^{\pi}=1^{e} \rightarrow 2J^{\pi}=1^{\circ},2^{\circ},3^{\circ}$.  Whereas for the excited metastable states then,
 $2J^{\pi}=3^{e} \rightarrow 2J^{\pi}=1^{\circ},3^{\circ},5^{\circ}$ and
 $2J^{\pi}=5^{e} \rightarrow 2J^{\pi}=3^{\circ},5^{\circ},7^{\circ}$,
 $2J^{\pi}=7^{e} \rightarrow 2J^{\pi}=5^{\circ},7^{\circ},9^{\circ}$,
$2J^{\pi}=9^{e} \rightarrow 2J^{\pi}=7^{\circ},9^{\circ},11^{\circ}$ are necessary.

%
%

\begin{table}
\caption{Comparison of the NIST~\cite{NIST2014} tabulated data with the present theoretical energies
obtained by using the GRASP code. Relative energies with respect to the ground state are given  in eV.
A sample of the 19 lowest NIST levels of the residual W$^{2+}$ ion are
compared with two different GRASP calculations,  449- and 573-level approximations.}
\label{tab1}.
\begin{tabular}{llcccccc}
\br
Level         &  CONFIG 	     	&  Term		     	& NIST  		 		&GRASP 	    	& GRASP  		    &$\Delta_1$		&$\Delta_2$\\
		&			     	&			   	& Energy$^{\dagger}$ 	&Energy$^a$	& Energy$^b$ 		   &Energy$^c$	 &Energy$^d$ \\	
		&			     	&			    	& (eV)		 		&(eV) 			& (eV) 			& (eV)		& (eV)\\	
\mr
 1  		& $5d^4$ 			&  $\rm^5D_{0}$     	& 0.000000			&0.000000	&    0.000000 		    &--			   &--  \\
 2  		& $5d^4$			&  $\rm^5D_{1}$     	& 0.279733			&0.166265	&    0.163990		    &-0.113		&-0.116\\
 3  		& $5d^4$			&  $\rm^5D_{2}$     	& 0.553117			&0.370834	&    0.355578		    &-0.182		&-0.187\\
 4  		& $5d^4$			&  $\rm^5D_{3}$     	& 0.778349			&0.574100	&    0.568446		    &-0.204		&-0.210\\
 5  		& $5d^4$			&  $\rm^5D_{4}$     	& 0.953027			&0.769428	&    0.762980		    &-0.184		&-0.190\\
 \\
 6  		& $5d^4$			&  $\rm^3P2_{0}$   	& 1.227977			&1.196092	&    1.183900		    &-0.032		&-0044\\
 7  		& $5d^4$			&  $\rm^3P2_{1}$   	& 1.597044			&0.931460	&    0.854302		    &-0.666		&-0.743\\
 8  		& $5d^4$			&  $\rm^3P2_{2}$   	& 2.060751			&1.095965	&    1.017557		    &-0.965		&-1.045\\
\\
  9  		& $5d^3(^4F)6s$	&  $\rm^5F_{1}$     	& 1.359926 	     		&1.513361	&    1.494708		   &+0.153		   &+0.135\\
10  		& $5d^3(^4F)6s$	&  $\rm^5F_{2}$     	& 1.540763	    		&1.783970	&    1.758645		   &+0.243		  &+0.218\\
11  		& $5d^3(^4F)6s$	&  $\rm^5F_{3}$     	& 1.864479	    		&1.343193	&    1.262595		   &-0.521		   &-0.602\\
12  		& $5d^3(^4F)6s$	&  $\rm^5F_{4}$     	& 2.154895	    		&1.627609	&    1.546138		   &-0.527		   &-0.608\\
13 		& $5d^3(^4F)6s$	&  $\rm^5F_{5}$     	& 2.434064	    		&1.916626	&    1.835504		   &-0.517		   &-0.598\\
\\
14  		& $5d^4$			&  $\rm^3F2_{2}$   	& 1.734804			&1.983106     	&    1.956663		   &+0.203		&+0.222\\
15  		& $5d^4$			&  $\rm^3F2_{3}$   	& 1.847340			&1.938382     	&    1.920567		   &+0.091		&+0.370\\
16  		& $5d^4$			&  $\rm^3F2_{4}$   	& 2.278383			&2.262686     	&    2.245308		   &-0.016		&-0.033\\
\\
17  		& $5d^4$			&  $\rm^3H_{4}$     	& 1.698701			&1.894375	&   1.890280		   &+0.196		&+0.192\\
18  		& $5d^4$			&  $\rm^3H_{5}$     	& 2.073417			&2.222915	&   2.217443		   &+0.149		&+0.139\\
19  		& $5d^4$			&  $\rm^3H_{6}$     	& 2.278941			&2.409139	&   2.403855		   &+0.130		&+0.125\\
\mr
\end{tabular}
\\
\begin{flushleft}
$^{\dagger}$Energies from the NIST Atomic Spectra Database  \cite{NIST2014}.\\
$^a$GRASP theoretical energies from the 449-level approximation.\\
$^b$GRASP theoretical energies from the 573-level approximation.\\
$^c\Delta_1$ energy difference (eV) of the 449-level approximation with NIST \cite{NIST2014} values.\\
$^d\Delta_2$ energy difference (eV) of the 573-level approximation with NIST \cite{NIST2014} values.\\
\end{flushleft}
\end{table}

For the ground and metastable initial states of the tungsten ions studied here, the outer region electron-ion collision problem was solved (in the resonance region below and  between all thresholds) using a  fine energy mesh of 10$^{-5}$ Rydbergs ($\approx$ 0.136 meV)  for the $5d^4 6s\,\, {\rm ^6D}_{1/2}$ ground level and 10$^{-4}$ Rydbergs ($\approx$ 1.36 meV) for the excited levels investigated. The $jj$-coupled Hamiltonian diagonal matrices were adjusted so that the theoretical term energies matched the recommended NIST values~\cite{NIST2014}. We note that this energy adjustment ensures better positioning of resonances relative to all thresholds included in
the calculation \cite{McLaughlin2012a,McLaughlin2012b}.

In the present work the DARC PI cross-section calculations for  Ta-like tungsten ions were convoluted with Gaussian profiles of 50 or 100 meV FWHM simulating the experimental photon energy bandwidths.

%
%
%
%
\begin{figure}
\begin{center}
\includegraphics[scale=1.0,width=13cm]{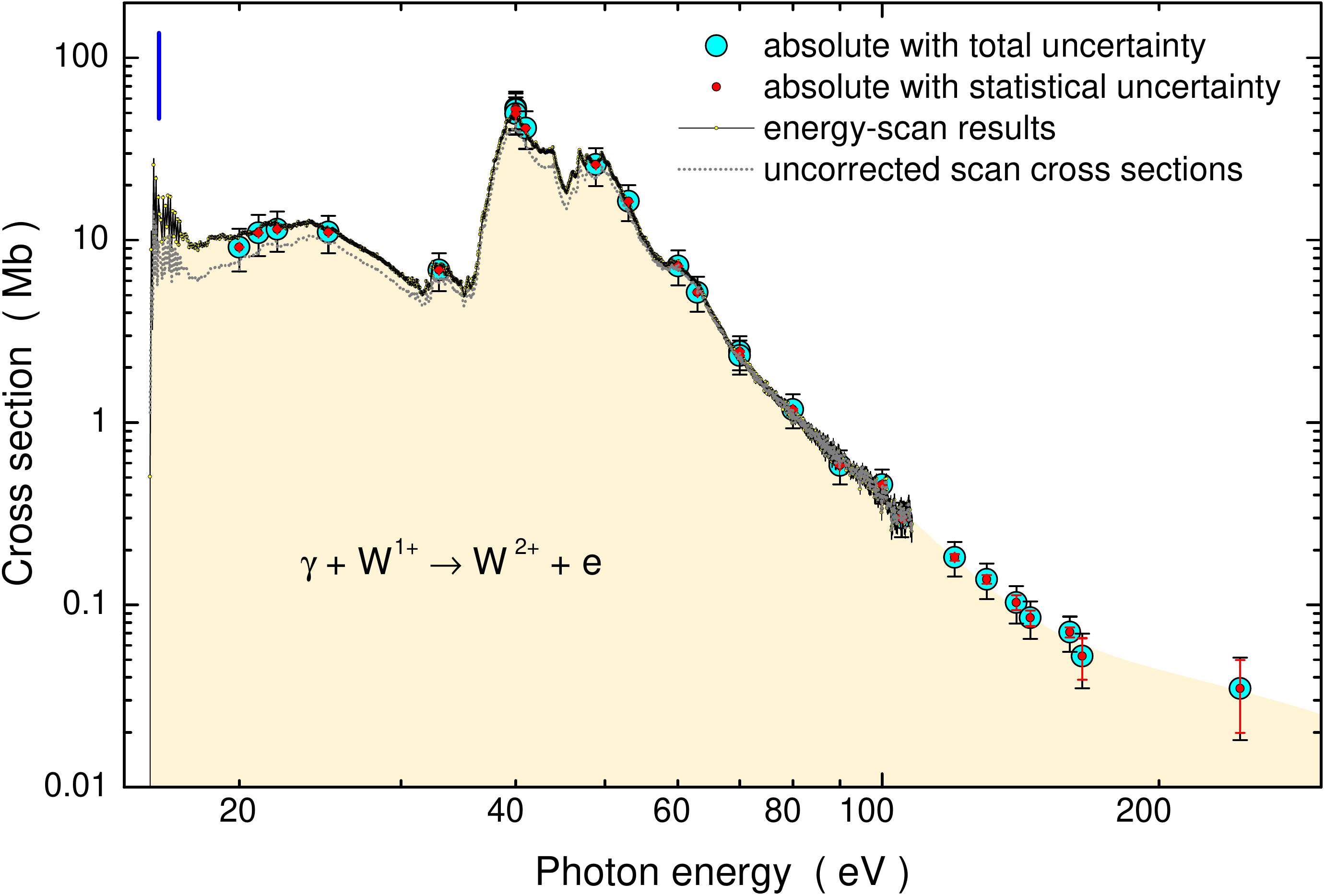}
\caption{\label{Fig:exp-overview} (Colour online) Photoionization of W$^{+}$ ions  measured at energy resolution 100~meV. Energy-scan measurements (small open circles with statistical error bars) were normalized to absolute cross-section data represented by large (cyan) shaded circles with total error bars and smaller full (red) circles with statistical uncertainties.	The (blue) vertical bar at 16.37~eV indicates the ground-state ionization potential. The data have been corrected for the effects of higher-order radiation (see text and appendix). The uncorrected energy scan is shown as a dotted (grey) line.}
\end{center}
\end{figure}

%
%
%
%
\begin{figure}
\begin{center}
\includegraphics[scale=1.0,width=8.3cm]{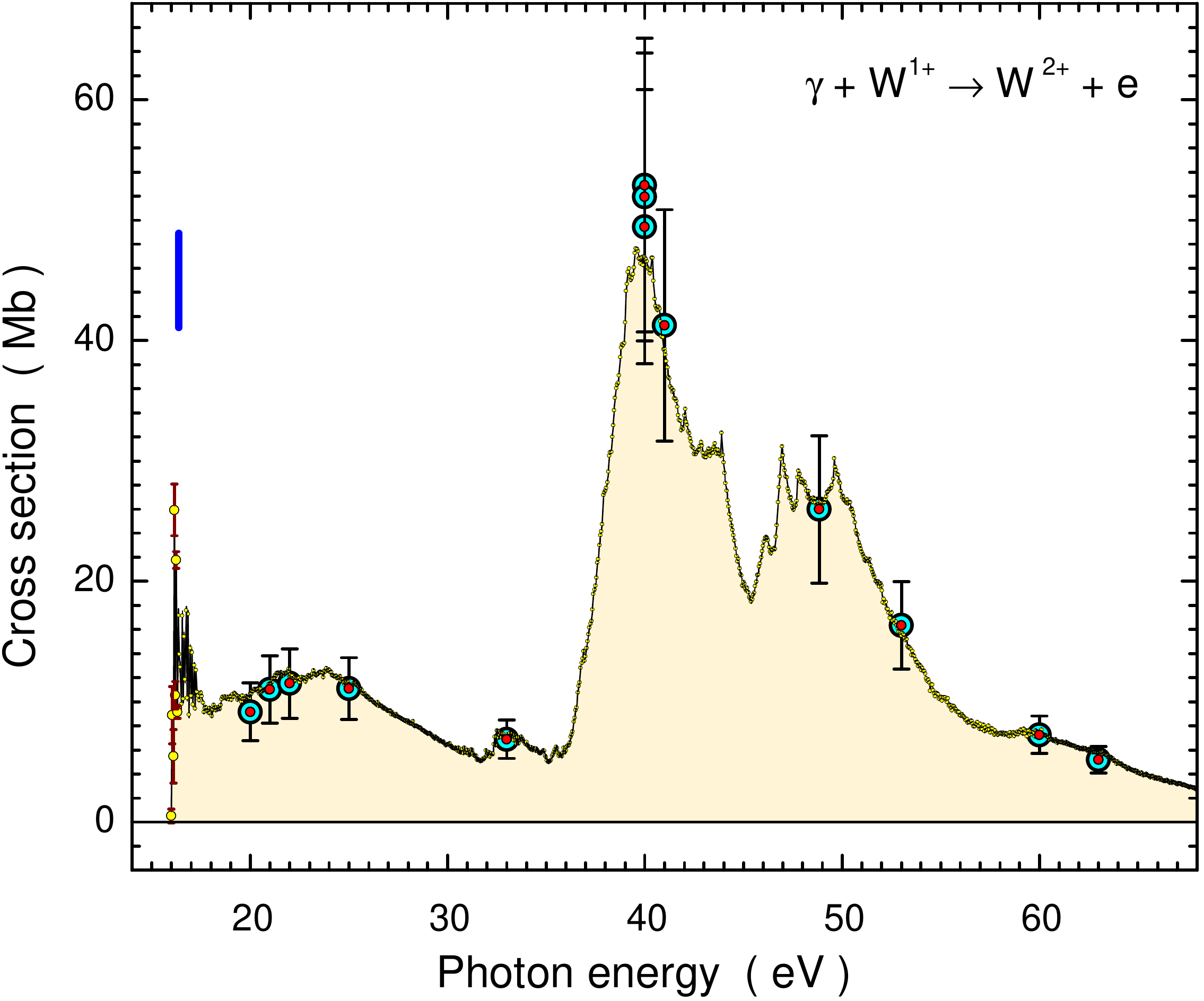}
\caption{\label{Fig:exp-detail}(Colour online) Detail from Fig.~\ref{Fig:exp-overview} with data shown on linear cross-section and energy scales. The symbols are the same as in Fig.~\ref{Fig:exp-overview}. The first seven lowest-energy data points of the energy scan are enlarged and their statistical error bars are emphasized by bold (brown) lines with endcaps. The statistical uncertainties of the absolute data points are negligibly small in the energy range of the figure. This is also true for the energy-scan data except for the range 16 to 17~eV. Total uncertainties of absolute data points are indicated by the solid error bars with large endcaps. The total relative error of the point at 16~eV is about 60\% due to the uncertainty
of the correction for higher-order effects (see text).
}
\end{center}
\end{figure}
%
%
%
%
\begin{figure}
\begin{center}
\includegraphics[scale=1.0,width=8.3cm]{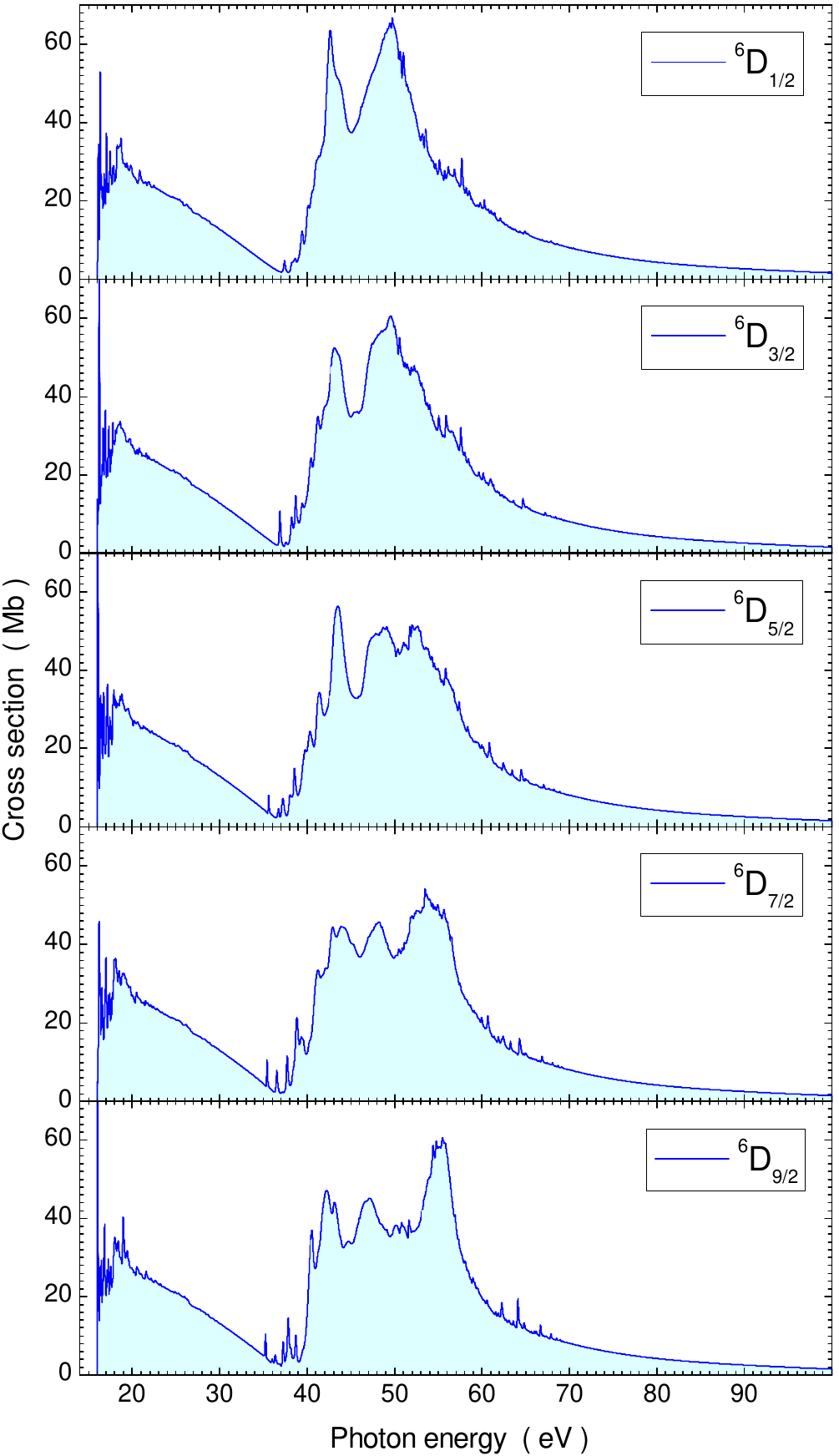}
\caption{\label{{Fig:theo-6D}}(Colour online) Theoretical photoionization cross sections from lowest-term
	W$^{+}(5d^4 6s\,\, {\rm ^6D}_J)$ ions with total angular momentum quantum numbers
	$J$~=~1/2, 3/2, 5/2, 7/2 and 9/2 individually specified in each panel. The theoretical data
	were obtained from 573-level DARC calculations and then convoluted with a 100~meV FWHM Gaussian profile.}
\end{center}
\end{figure}

\section{Results and Discussion}
\label{Sec:Results}

Fig.~\ref{Fig:exp-overview} presents the measured cross section for single photoionization of W$^+$ ions on a double-logarithmic scale. The measurements have been corrected for the effects of higher-order radiation as described in the appendix. For comparison, also the uncorrected energy scan is displayed. The difference between the corrected and uncorrected scan data is within the total error bars of the experiment. In the energy range investigated, 16 to 245~eV, the cross section spans more than three orders of magnitude in size. It is dominated by very broad features with relatively small and narrow resonances occurring in certain energy ranges. The energy-scan data are shown by small open circles with error bars which are mostly so small that they can only be seen between 16 and at most 18~eV and again at energies above 90~eV where the signal count
rates were small. The absolute cross sections are shown twice, once as small solid (red) points with statistical error bars and once as large (cyan-)shaded circles with their total absolute uncertainties. At 245~eV the statistical uncertainty is the dominating source of possible error.

The most interesting features in the photoionization cross section of W$^+$ are found in the energy range from 16 to about 70~eV. For better display of the cross section details Fig.~\ref{Fig:exp-detail} highlights the interesting energy range and shows the experimental data on linear cross-section and photon-energy scales.

The apparent onset of the cross section is near 16~eV which is close to the ground-level ionization potential. It should be mentioned, however, that below 16~eV no measurements were possible. Although the cross section almost goes to zero at 16~eV within the large uncertainties of these low-energy measurements there might still be a sizable cross section contribution below 16~eV which would arise from some of the many metastable levels that are within reach of the ion source. Towards higher energies a strong increase of the cross section above 36~eV indicates the onset of  new ionization channels beyond the removal of a $6s$ or $5d$ electron. We assign the first big peak in the cross section to excitation and ionization of a $4f$ electron. The next strong peak with its onset at around 45 eV is attributed
to the opening of the $5p$ subshell.

Similarly strong peaks are predicted by theory for photoionization of W$^+$ ions from all the initial levels investigated. The theoretical results for the ten energetically lowest initial levels are shown in the next three graphs. Fig.~\ref{{Fig:theo-6D}} illustrates the photoionization results for each individual fine-structure component within the term $5d^4 6s\,\, {\rm ^6D}$ with total angular momentum quantum numbers $J$~=~1/2, 3/2, 5/2, 7/2 and 9/2. Fig.~\ref{{Fig:theo-6S}} displays the photoionization cross section for the only level within the $5d^5\,\,{\rm ^6S}$ term with $J$~=~5/2. Finally,  Fig.~\ref{{Fig:theo-4F}} shows the theoretical results for all levels within the term $5d^3 6s^2\,\, ^4F$ with $J$~=~3/2, 5/2, 7/2 and 9/2. All theoretical cross sections were convoluted with a
Gaussian of 100~meV full width at half maximum in order to simulate the experimental conditions of the data displayed in Fig.~\ref{Fig:exp-overview}.

The excitation energies of the excited $5d^4 6s\,\, {\rm ^6D}_J$ levels investigated in Fig.~\ref{{Fig:theo-6D}} are 0.188~eV for $J$~=~3/2,  0.393~eV for $J$~=~5/2,  0.585~eV for $J$~=~7/2 and 0.762~eV for $J$~=~9/2~\cite{NIST2014}. At energies right above the associated ionization thresholds and below 18~eV theory predicts sharp resonance features. With increasing photon energy a region of smooth energy dependence is predicted up to about 35~eV where narrow resonances start again to appear. The smoothly decreasing cross section arises from direct photoionization mainly of one of the $5d$ electrons. The narrow resonances occurring predominantly in the photoionization of the ${\rm ^6D}_{9/2}$ level are most likely due to $4f$ excitations which add up to a strong increase of the cross section above 35~eV while producing broad peak features. It is worth noting that smooth broad cross section peaks are also seen in the theoretical results before convolution. Apparently, the extreme density of excited states accessible by excitation of a $4f$ electron in W$^+$ leads to overlapping and interacting resonances similar to the behaviour found recently in photorecombination of tungsten ions in charge states around $q=20$~\cite{Mueller2015b,Schippers2011b,Spruck2014}. At higher photon energies the onset of $5p$ excitation and then also direct ionization of the $5p$ subshell is expected. However, the signature for opening the $5p$ subshell is not clearly evident from the theoretical results displayed in Fig.~\ref{{Fig:theo-6D}}. Beyond 55 to 60~eV the calculated cross sections rapidly drop from a level of 60~Mb at 55~eV to about 2~Mb at 100~eV. Qualitatively, the theoretical cross sections shown in Figs.~\ref{{Fig:theo-6D}} and \ref{{Fig:theo-6S}} display a similar overall behaviour.
%
%
%
%
\begin{figure}
\begin{center}
\includegraphics[scale=1.0,width=8.3cm]{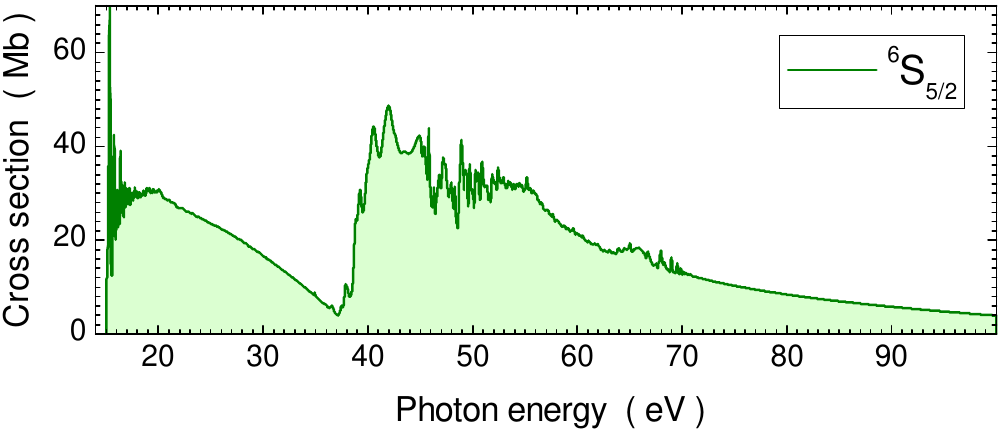}
\caption{\label{{Fig:theo-6S}}(Colour online) Theoretical photoionization cross sections of second lowest-term W$^{+}(5d^5\,\,{\rm ^6S_{5/2}})$ ions. The theoretical data were obtained from 573-level DARC calculations and then convoluted with a 100~meV FWHM Gaussian profile.}
\end{center}
\end{figure}
%
%
%
%
\begin{figure}
\begin{center}
\includegraphics[scale=1.0,width=8.3cm]{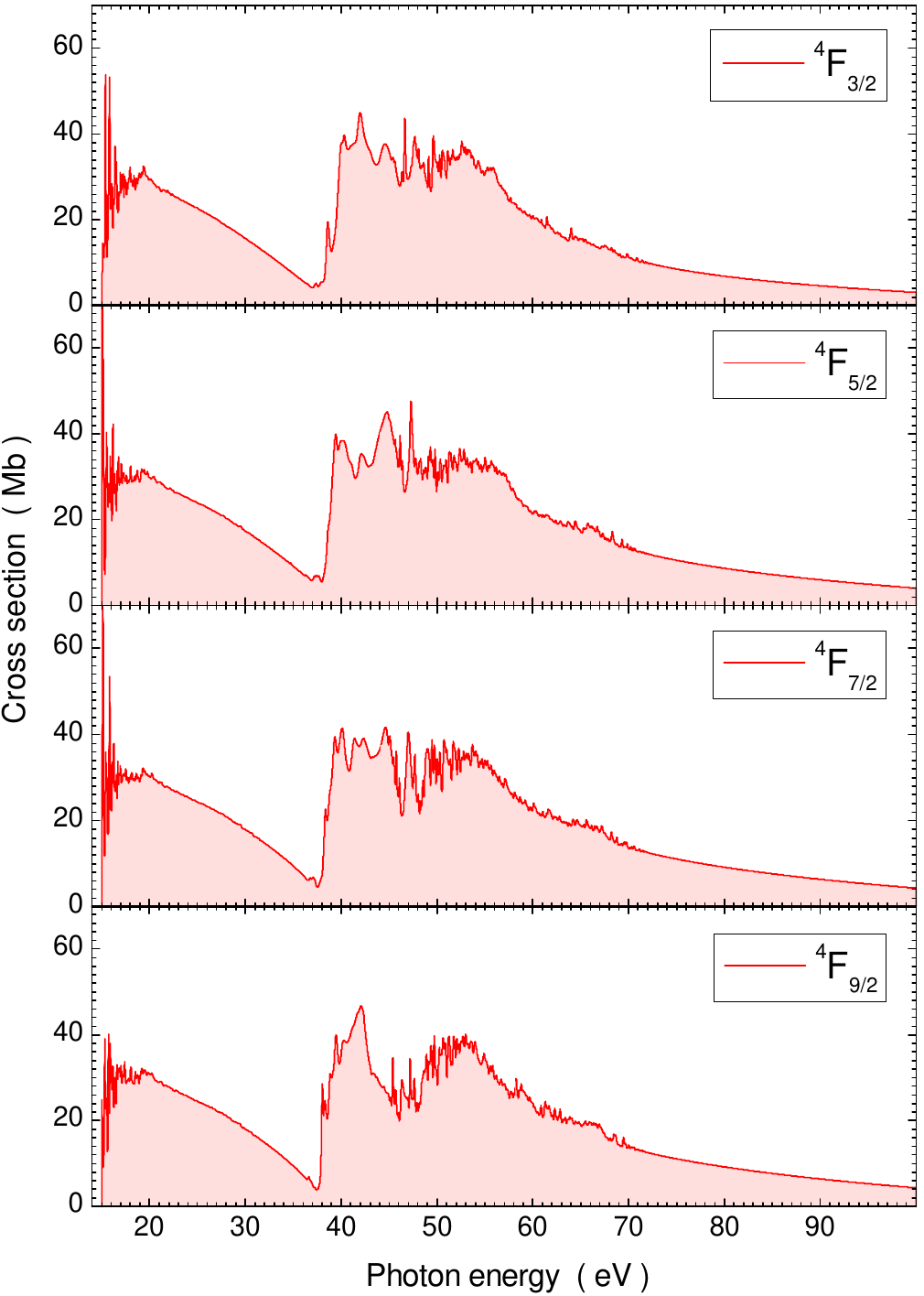}
\caption{\label{{Fig:theo-4F}}(Colour online) Theoretical photoionization cross sections of $^4$F - term W$^{+}(5d^3 6s^2\,\, ^4F_{J})$ ions  with total angular momentum quantum numbers $J$ = 3/2, 5/2, 7/2 and 9/2 individually specified in each panel. The theoretical data were obtained from 573-level DARC calculations and then convoluted with a 100~meV FWHM Gaussian profile.}
\end{center}
\end{figure}

Fig.~\ref{{Fig:theo-6S}} illustrates the calculated photoionization cross section of the lowest term, ${\rm ^6S}$, of the first excited configuration ($5d^5$) comprising only one level with the spectroscopic notation ${\rm ^6S}_{5/2}$. Its excitation energy from the ground level is 0.920~eV~\cite{NIST2014}. Slightly different from the $5d^4 6s\,\, {\rm ^6D}_J$ levels in the ground configuration more structure is predicted in the cross section between 45 and 55~eV. Also the decrease towards higher photon energies is not equally steep.

The predictions of the photoionization cross sections from the next higher levels in W$^+$, associated with the $5d^3 6s^2\,\, {\rm ^4F}$ term, are displayed in Fig.~\ref{{Fig:theo-4F}}. The excitation energies above the ground level of W$^+$ are 1.080~eV for $J$~=~3/2, 1.401~eV for $J$~=~5/2 and 1.663~eV for $J$~=~7/2~\cite{NIST2014}. The $5d^3 6s^2\,\, {\rm ^4F_{9/2}}$ level  is not listed with this notation in the NIST database. This is attributed to the increasing importance of mixing effects to be considered and the associated difficulty to unambiguously assign a level notation.  The present calculations give 1.318~eV for $J$~=~9/2. The calculated spectra for the ${\rm ^4F_{J}}$ levels are very similar to the photoionization cross section of W$^+({\rm ^6S}_{5/2})$ shown in Fig.~\ref{{Fig:theo-6S}}.

All ten calculated cross sections show the same most prominent features. Narrow resonances at energies up to about 18~eV are followed by a smoothly decreasing function of energy which can be associated with $6s$ and $5d$ outer-shell ionization. In all cases a steep increase of the cross section is caused by the opening of the $4f$ subshell and the onset of vacancy production in the $5p$ subshell produces a further strong peak feature above 45~eV. At energies beyond about 55~eV all cross sections decrease with increasing photon energy. They all show indications of additional small structures between 65 and 70~eV and a smooth dependence up to 125~eV (where the range between 100 and 125~eV is not shown in the figures).
 %
%
%
%
\begin{figure}
\begin{center}
\includegraphics[scale=1.0,width=8.3cm]{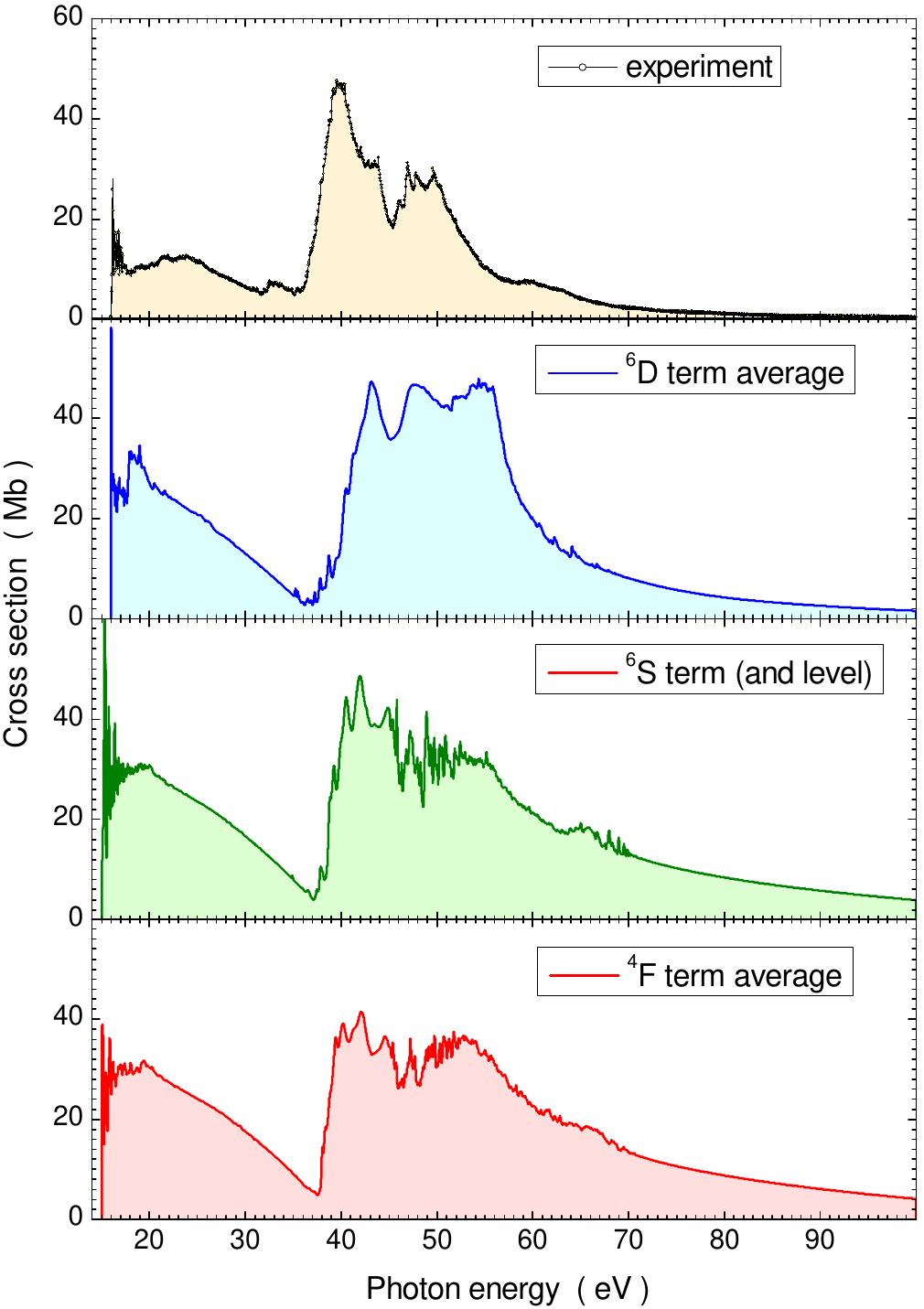}
\caption{\label{{Fig:theo-exp-all}}(Colour online) Comparison of experimental and term-averaged theoretical photoionization cross sections of W$^{+}$ ions at 100~meV energy resolution. The three lower panels show the theoretical results obtained from 573-level DARC calculations for photoionization from the energetically lowest terms $5d^4 6s\,\ {\rm ^6D}$, $5d^5\,\ {\rm ^6S}$ and $5d^3 6s^2\,\ {\rm ^4F}$, respectively.}
\end{center}
\end{figure}

The interpretation of the experimental results shown in Fig.~\ref{Fig:exp-overview} is complicated by the possible presence of long-lived excited levels in the primary ion beam used for the cross section measurements. Since photon energies below 16~eV were not accessible by the experimental setup it was not possible to draw final conclusions on beam fractions of those ions from the observation of photoionization signal below the ground-level ionization potential. With the measured cross section being as little as (0.5$\pm$0.6)~Mb  at 16.0~eV and already (25.9$\pm$6.8)~Mb  at 16.15~eV there is an indication, however, that a substantial fraction of the ion beam must have been associated with the lowest-energy term of W$^+$. According to the NIST level energies~\cite{NIST2014} mentioned above, the ${\rm ^6D}$ term averaged ionization potential is (15.86$\pm$0.15)~eV which is compatible with the photoionization onset observed in the present experiment. The term-averaged apparent ionization onsets in the present calculations for the lowest three terms in W$^+$ are 16.0~eV for the ${\rm ^6D}$ term and 15.0~eV for both the ${\rm ^6S}$ and ${\rm ^4F}$ terms. So the experiment is in agreement with the calculation for the ${\rm ^6D}$ term. The fact, that the experimental cross section is so small at 16.0~eV even when considering the very large error bar, may indicate that possible contributions from the first excited terms are well below 10\%. This observation can be discussed in the context of  lifetimes of metastable levels and  time delay between the production of an (excited) ion and its arrival at the photon-ion interaction region.

The time of flight of 6-keV $^{186}$W$^+$ ions for the 486-cm path from the ECR ion source to the center of the photon-ion interaction region is readily determined to be about 62~$\mu$s. An estimate of decay probabilities (equivalent to partial lifetimes) of the dipole-forbidden transitions between all levels within the $5d^4 6s$, $5d^3 6s^2$ and $5d^5$ configurations, all with even parity, has been obtained by employing the GRASP code. Ten configurations ($5d^4 6s$, $5d^4 6p$, $5d^4 6d$, $5d^3 6s^2$, $5d^3 6p^2$, $5d^3 6d^2$, $5d^5$, $5d^3 6s 6p$, $5d^3 6s 6d$, and $5d^3 6p 6d$) were used as a basis for representing the structure of W$^+$ (see section~\ref{sec:Theory}). The results  show that even the fastest E2 transitions have partial lifetimes exceeding the ions' time of flight by more than a factor of three. The less probable M1 transitions have lifetimes of even more than 0.1 s. As a result of these comparisons one would expect that levels within the $5d^4 6s$, $5d^3 6s^2$ and $5d^5$ configurations almost all survive the flight time of the excited ions after their extraction from the ion source. However, beside their flight time there is also a drift time of the ions before they leave the source volume. This has been observed previously~\cite{Borovik2009a} for an unambiguous case, the metastable $1s2s\,\ {\rm ^1S}$ level in heliumlike Li$^+$ with a known lifetime of about 0.5~ms~\cite{Saghiri1999}. In electron-impact ionization experiments with Li$^+$ ions Borovik \etal~\cite{Borovik2009a} found no evidence for the presence of Li$^+(1s2s\,\ {\rm ^1S}$) in the parent ion beam although the ion flight time was only about 8~$\mu$s, i.e., over 60 times less than the lifetime of the $1s2s\,\ {\rm ^1S}$ level. The conclusion was that the ions spend a much longer time drifting within the ion source between production and extraction  than traveling from the ion source to the interaction region, giving them enough time to even let the $1s2s\,\ {\rm ^1S}$ level decay in spite of its 0.5~ms lifetime. This interpretation was also supported by other ionization experiments with a focus on metastable $1s2s$ heliumlike ions in which the same type of ion source was employed~\cite{Renwick2009,Mueller2014g}. On the basis of such previous observations, one may speculate that  in the present photoionization experiment a considerable fraction of the long-lived levels in the lowest configurations of W$^+$ may have decayed before the ions reached the photon-ion interaction region - in accord with the findings reported in the previous paragraph.

Clearly, one cannot expect that only the lowest level within a given term is populated in the ion-source plasma. The electron energy distribution in an electron-cyclotron-resonance heated plasma in which also multiply charged ions can be produced must be expected to contain components with at least several tens of eV. Compared to this high energy, the fine-structure energy splitting within a $J$-multiplet is negligibly small. The only reasonable assumption about the population of the fine-structure levels of W$^+$ ions within a given term is that of a statistical distribution with the statistical weights given by $2J+1$. Calculated lifetimes of the levels within the ground term are between several days and a few years. Therefore, comparison of experimental cross section data with term-averaged theoretical cross sections is the most meaningful.

Fig.~\ref{{Fig:theo-exp-all}} thus shows the experimental data from Fig.~\ref{Fig:exp-overview} in comparison with the term-averaged theoretical results for the $5d^4 6s\,\ {\rm ^6D}$, $5d^5\,\ {\rm ^6S}$ and $5d^3 6s^2\,\ {\rm ^4F}$ terms. If W$^+$ ions in a given level are present in the parent ion one has to assume that all other levels belonging to that same multiplet are also present and that they are statistically populated. Hence, the remaining question is which multiplets or terms are populated in the ion source and survive the time of flight of the ions from the ion source to the photon-ion interaction region. As discussed above, there are at least 118 excited levels in the first three lowest-energy configurations that are expected to be metastable since they all have the same (even) parity. In principle, all these levels might have contributed to the experimental result (together with the $5d^4 6s\,\ {\rm ^6D_{1/2}}$ ground level). To perform cross-section calculations for 109(=119-10) more levels of the W$^+$ ion would require an enormous computational effort and dedication of resources. Thus, the experiment can only be compared to a limited set of theoretical data. In spite of this limitation one can state a number of observations of theory describing prominent features in the experiment.
 %
%
%
%
\begin{figure}
\begin{center}
\includegraphics[scale=1.0,width=8.3cm]{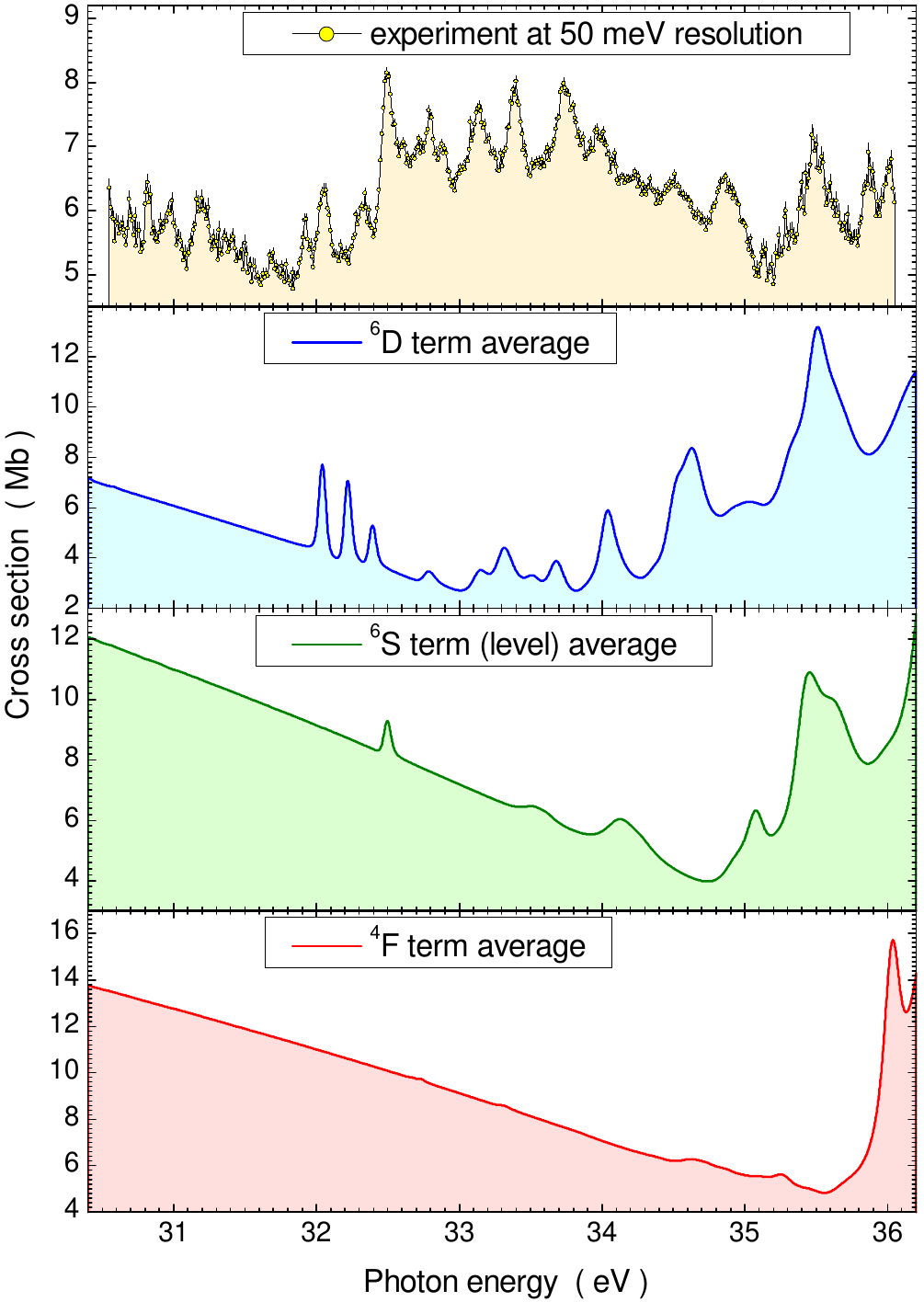}
\caption{\label{{Fig:theo-exp-resonances}}(Colour online) Comparison of experimental and term-averaged theoretical photoionization cross sections of W$^{+}$ ions at 50~meV energy resolution in an energy range where narrow resonances occur in the cross section. The three lower panels
show the theoretical results obtained from 573-level DARC calculations for the energetically lowest terms $5d^4 6s\,\ {\rm ^6D}$, $5d^5\,\ {\rm ^6S}$ and $5d^3 6s^2\,\ {\rm ^4F}$, respectively. 	The theoretical spectra are shifted in energy by -3.2~eV, -2.5~eV and -2.1~eV, respectively. For more details see text.}
\end{center}
\end{figure}

Theory predicts narrow resonances at energies up to about 18~eV. Although the step width is too coarse in the experiment there is clear indication of rapid oscillations in the cross section at low photon energies. A relatively smooth energy dependence of the cross section follows at increasing photon energies. The experimental cross section goes over a very broad maximum while theory predicts a monotonically decreasing cross section. Around 35~eV the experimental result shows some structure  and also narrow resonances again. The only channel calculated to provide significant amounts of resonance structure in  this energy region is that for the ${\rm ^6D}$ term. This is further evidence of  a substantial fraction of ions in their lowest-energy term present in the experiment.

All calculations show the rapid increase of the cross section at photon energies above 35~eV that also characterizes the experimental result. The cross section maximum reached in each of the calculations is in close proximity of the experimental maximum. In the energy range 40 to 55~eV the details of the experimental cross section structure are not closely reproduced by the calculations or a reasonable combination of contributions from the investigated terms. These differences in the details are ascribed to the still very limited basis set of the calculations which was chosen to keep the computational effort managable.

Beyond 55~eV the experimental cross section drops off rapidly. A similarly rapid decrease is only seen in the calculations for the ${\rm ^6D}$ term. The bump at about 60~eV in the experimental cross section is also seen in the theoretical data for the ${\rm ^6S}$ and ${\rm ^4F}$ terms. At energies beyond 70~eV theory overestimates the experimental single ionization cross section. Parts of the calculated ionization contributions may in fact end up in multiple-ionization channels after relaxation of the photoionized intermediate state formed by the removal of a single electron from W$^+$.

In energy ranges where narrow resonances could be observed, additional energy scans of the cross section were measured at 50~meV resolution. The most prominent occurrence of narrow features is in the energy range between about 30 and 36~eV. The top panel of Fig.~\ref{{Fig:theo-exp-resonances}} shows energy-scan results normalized to the absolute cross sections shown in Fig.~\ref{Fig:exp-overview}. Detailed resonance structure can be seen with the strongest peak feature occurring at about 35.5~eV just before the steep rise in the cross section due to the opening of the $4f$ subshell. The associated energy ranges of features in the theoretical cross sections are shown in the three lower panels. The energy axes of the calculated data were shifted in order to match certain features in the experimental cross section. It was felt that the experimental peak at 35.5~eV might correspond to the relatively broad resonance structures in the ${\rm ^6D}$ and ${\rm ^6S}$ calculations occurring just below the steep rise in the cross section. Therefore, the theoretical energy scales were adjusted by -3.3 and -2.5~eV, respectively. In the ${\rm ^4F}$ calculations no corresponding peak could be found. The energy axis of the ${\rm ^4F}$ spectrum was shifted by -2.1~eV to match the steep onset of the $4f$ contribution to the cross section. Again it is the calculation for the ${\rm ^6D}$ initial term of W$^+$ ions that matches best with the experiment although the fine details of the measurements are not reproduced by theory.

Although, the overall shapes of all the calculated cross section curves are very similar to the experimental results it is obvious that a conclusive comparison between theory and experiment cannot easily be made. Over much of the displayed photon energy range, the calculated theoretical cross sections are consistently larger than the experimental cross section by factors 2--3 or more (in particular, at higher energies). A previous isolated case,  the photoionisation of singly-charged selenium~\cite{McLaughlin2012b} also revealed large discrepancies between experiment and theory, however this must be viewed in the context of Ar$^+$ , Kr$^+$ and Xe$^+$ results, which have similar initial configurations and exhibit very good agreement between theory and experiment. Converging the theoretical wavefunction through additional configuration interaction for such a complicated species is a slow process, and  although we believe our model is reasonable,  it may require revisiting when future computational resources become available. A different theoretical approach considering cases with very large numbers of contributing excited levels has recently been suggested~\cite{Flambaum2015}, but  has not yet been applied to a specific case.

\section{Summary and Conclusions}\label{sec:Conclusions}
Experimental and theoretical photoionization cross sections for W$^+$ ions are presented. The experimental cross sections were measured on an absolute scale employing the photon-ion merged-beam facility at the Advanced Light Source. The theoretical data were obtained from large-scale close-coupling calculations within the Dirac-Coulomb R-matrix approximation (DARC). The comparison of the measured and calculated results is complicated by the possible presence of long-lived excited states in the parent ion beams used for the experiments. More detailed modeling of the experimental data by theory would require calculations for at least all the 119 levels in the lowest configurations of the W$^+$ ion which is presently beyond the availability of computer resources. There are indications, though,  in the measured cross section that most of the parent ions were in the ground-state ${\rm ^6D}$ term. Given the existing limitations and considering the complexity of Ta-like tungsten with its open $5d$ subshell, one can conclude that the
main features of the experimental results are reasonably well reproduced by the theoretical calculations. This result for a complex singly charged ion where strong electron-electron correlation effects are to be expected  is encouraging for applying a similar theoretical approach to other more highly charged tungsten ions where the relative importance of electron-electron interactions is reduced.

\ack
We acknowledge support by Deutsche Forschungsgemeinschaft
under project number Mu-1068/20 in addition to grants from the US Department
of Energy (DOE) under contracts DE-AC03-76SF-00098 and  DE-FG02-03ER15424.
C P Ballance was supported by NASA and NSF grants  through Auburn University.
B M McLaughlin acknowledges support by the US National Science Foundation through a grant to ITAMP
at the Harvard-Smithsonian Center for Astrophysics, Queen's University Belfast
for the award of a visiting research fellowship (VRF) and the hospitality of AM, SS
and the University of Giessen during a recent  visit.
The computational work was carried out at the National Energy Research Scientific
Computing Center in Oakland, CA, USA and at the High Performance
Computing Center Stuttgart (HLRS) of the University of Stuttgart, Stuttgart, Germany.
This research also used resources of the Oak Ridge Leadership Computing Facility
at the Oak Ridge National Laboratory, which is supported by the Office of Science
of the U.S. Department of Energy under Contract No. DE-AC05-00OR22725.
The Advanced Light Source  is supported by the Director, Office of Science, Office of Basic Energy Sciences,
of the US Department of Energy under Contract No. DE-AC02-05CH11231.

\appendix
\section*{Appendix}
\setcounter{section}{1}
It is well known that synchrotron light is usually accompanied by higher-order radiation. Transverse oscillations of electrons in an undulator insertion device introduce higher harmonics into the motion which give rise to radiation with wavelengths $\lambda_n = \lambda_1 /n$ where $\lambda_1$ is the wavelength of the fundamental~\cite{Attwood1999}. In particular, spherical grating monochromators (SGM)  disperse these higher-$n$ harmonics in $n^{th}$ order and due to the efficiency of the grating for the 1$^{st}$ and higher order radiation, there is a non-zero contribution to the photon beam on the optical axis of the monochromator. Higher harmonic radiation has characteristic angular patterns. The even harmonics radiate with angular distributions that have zero density at zero angle and peak at a finite (but very small) angle. The odd harmonics peak on axis and have relatively high brightness. The grating and all optical elements of the monochromator transporting the radiation to the experimental station have an influence on the mixture of harmonics that is present in an experiment, with design specifications tailored to minimize the higher order components delivered to the target.

Higher-order components of the photon beam are often used for energy-calibration purposes  over wide energy ranges.
At the expense of intensity they can be used to get information about cross section features whose energy is beyond the available range of the first harmonic. While there are ways of making positive use of the presence of higher-order radiation, it has  adverse effects on the measurement of absolute photoionization cross sections. Such effects have to be considered and corrections made to the measurements.

\subsection{Effects of higher-order radiation components in the photon beam}
As discussed in Sec.~\ref{sec:exp} merged-beam photoionization cross sections for ions are determined from
measurements by making use of Eq.~\ref{Eq:xsec} which is based on the assumption of a monoenergetic photon beam.
When higher-order light is present in the photon beam the measured count rate of photo-ions consists of several contributions. In principle, fractions of $n^{th}$ order radiation with $n=2, 3, 4,...$ are possible in a beam of predominantly 1$^{st}$ order light. This is especially the case at the IPB endstation of beamline 10 at the ALS when the first (the low-energy) spherical grating  is used because the first-order efficiency drops significantly at low photon energies.  The presence of different photon-beam components produces a count rate of photo-ions
\begin{equation}
\label{Eq:rate}
R(E_\gamma)=\frac{I_{i}  \eta \mathfrak{F}(E_\gamma)}{q e^2 v_{i}}\sum_n {\frac{\sigma(nE_\gamma)I_{\gamma}^{(n)}}{Q(nE_\gamma)}}.
\end{equation}
The photodiode conversion efficiencies $Q(nE_\gamma)$ denote the numbers of electrons provided by the photodiode per
incident  $n^{th}$ order photon of energy $nE_\gamma$. The measured total photon-induced diode current is
\begin{equation}
\label{Eq:photocurrent}
I_{\gamma} = \sum_n {I_{\gamma}^{(n)}}
\end{equation}
which is a sum of individual currents induced by the $n^{th}$ order fractions of photons in the beam
\begin{equation}
\label{Eq:photocurrentN}
I_{\gamma}^{(n)}= \dot{N}_{\gamma}^{(n)} Q(nE) e\\
\end{equation}
with $\dot{N}_{\gamma}^{(n)}$ the number of $n^{th}$-order photons per second. Defining the fractions
$f_n(E_\gamma)$ of $n^{th}$-order photons in the incident photon beam one can  rewrite the individual photon flux components as
\begin{equation}
\dot{N}_{\gamma}^{(n)} = f_n \dot{N}_{\gamma}
\end{equation}
with the total number of photons per unit time
\begin{equation}
\dot{N}_{\gamma} = \sum_n {\dot{N}_{\gamma}^{(n)}}.
\end{equation}
When a photoionization cross section is measured, one first does not consider the higher-order fractions of photons present and therefore an apparent cross section $\sigma_{app}$ is determined which needs correction later on. The measured count rate of photoionized ions is associated with $\sigma_{app}$ via
\begin{equation}
\label{Eq:defsigapp}
R(E_\gamma)=\frac{I_{i}  \eta \mathfrak{F}(E_\gamma) I_{\gamma}}{q e^2 v_{i} Q(E_\gamma)} \sigma_{app}.
\end{equation}
The photo-ion rate $R(E_\gamma)$ of Eq.~\ref{Eq:defsigapp} is identical with the rate expressed by Eq.~\ref{Eq:rate}
provided the $n^{th}$ order components of the photon beam have identical beam profiles. Assuming that this is the case, $\sigma_{app}$ can be determined to be
\begin{equation}
\label{Eq:sigappfin}
\sigma_{app}(E_\gamma) = \frac{\sum_n {\sigma(nE_\gamma) f_n(E_\gamma)}}{\sum_n {Q(nE_\gamma) f_n(E_\gamma)}} Q(E_\gamma).
\end{equation}
There are two effects on the measured cross sections due to  the presence of $n^{th}$-order radiation in the photon beam:\\
(i) the apparent cross section $\sigma_{app}(E_\gamma)$ includes contributions from different energies $n E_\gamma$;\\
(ii) the apparent cross section $\sigma_{app}$ has a problem with normalization to the photon flux. With the presence of higher
order radiation the assumption of the conversion efficiency being determined by $Q(E)$ rather than a weighted sum of $Q(nE_\gamma)$
makes the resulting cross section $\sigma_{app}(E_\gamma)$ deviate from $\sigma(E_\gamma)$.\\
To obtain $\sigma(E_\gamma)$ from the measured $\sigma_{app}(E_\gamma)$ the fractions $f_n(E_\gamma)$ of $n^{th}$
order radiation have to be known. These fractions most often have to be determined by separate experiments.

\subsection{Assessment of relative fractions of higher-order radiation}

Possible procedures to determine fractions of higher-order radiation in a photon beam are the measurement of photoelectron energies
and the identification of characteristic photoionization cross-section features found in first order at photon energy $E_\gamma^{(1)}$ and
then again in $n^{th}$ order at energies  $E_\gamma^{(1)}/n $ with $n = 2, 3, 4,...$. Both procedures have been employed previously to correct
cross sections measured at the IPB endstation of beamline 10 at the ALS (see for example Refs.~\cite{Lu2006a,Esteves2011}).
For the correction of the present W$^+$ photoionization cross section measurement the latter procedure was
 employed using a number of different ions and taking advantage of the existing capability of the IPB endstation
 for absolute cross section measurement. The following cross section measurements were carried out:\\
(i) photoionization of He$^+$ with observation of the ionization threshold at 54.4~eV in first, second and third order;\\
(ii) photoionization of Xe$^{7+}$ with observation of the dominant broad $4d^9 5s 5f\,\, {\rm  ^2P^o}$ resonance at 122.1~eV~\cite{Mueller2014b}
in first, second and third order;\\
(iii) photoionization of Xe$^{3+}$ with observation of strong $4d \to 4f$ excitations centered at 87.0~eV~\cite{Emmons2005a} in first, second and third order;\\
(iv) photoionization of Xe$^{+}$ with observation of the strong $4d \to 5p$ excitations at energies between 55 and 57.5~eV~\cite{Andersen2001b} in first, second and third order;\\
(v) photoionization of Si$^{2+}$ with observation of the strong $2p \to nl$ excitations at energies between 110 and 135~eV~\cite{Mosnier2003a} in all orders up to the sixth.\\
For the measurements, following the typical procedures at the IPB endstation, the photon beam size was limited by horizontal and vertical baffles in order to cut off beam halos. By that,  about 20\% of the total available photon flux at a given bandwidth is lost. Changing the bandwidth does not have a noticeable influence on higher-order fractions. By cutting the photon beam with the baffles to half of its original intensity the second order fraction could be significantly reduced (by 35\%) as one would expect for higher even-order radiation whose intensity is peaked off axis. Since photon-ion experiments are photon-hungry due to the low particle densities of ion beams, typical experiments (and the measurements on W$^+$ included) are carried out using almost the full photon beam - just with halos cut off.

Cross section ratios $r_n = \sigma_{app}(E_\gamma/n) / \sigma_{app}(E_\gamma)$ for $n^{th}$ order cross section contributions at a given photon energy $E_\gamma/n$ relative to $1^{st}$ order measurements at the associated photon energy $E_\gamma$ are readily obtained from the above experiments. The measurements with Si$^{2+}$ ions show ratios $r_1:r_2:r_3:r_4:r_5:r_6 = 100:1.06:2.64:1.39:0.70:0.40$. With the exception of an increase between $r_2$ and $r_3$ the sequence appears to indicate that higher order fractions substantially decrease with $n$. By assuming that the fractions with $n \geq 4$ can be neglected and that the fractions $f_2$ and $f_3$ (and hence also $f_1$) are smooth
functions of $E_\gamma$ one finds that $f_2$ is a bell shaped function of $E_\gamma$ with a maximum of approximately 3\% at about 45~eV and only 1\% at 27 and 66~eV. Consistent with previous observation~\cite{Esteves2011} the dominant fraction $f_3$ is about 6\% at 20~eV but drops off more slowly with increasing photon energy than previously assumed. Instead of being negligible  at energies beyond 30~eV  $f_3$ is consistently found both from the measurements (ii) and (v) to be still about 3\% near 40~eV.

\subsection{Correction of cross section measurements for higher-order radiation effects}
From Eq.~\ref{Eq:sigappfin} one can formally derive the true cross section $\sigma(E_\gamma)$ from the measured apparent cross section to be
\begin{eqnarray}
\label{Eq:sigreal}
\sigma(E_\gamma) = \frac{1}{f_1(E_\gamma)} \times \\ \nonumber
 \times \left[\sigma_{app}(E_\gamma) \frac{\sum_{n=1}^{n_{max}} {Q(nE_\gamma) f_n(E_\gamma)}}{Q(E_\gamma)}
- \sum_{n=2}^{n_{max}} {\sigma(nE_\gamma) f_n(E_\gamma)} \right]
\end{eqnarray}
where $n_{max}$ is the index of the highest order of radiation to be considered. The first term on the right side of Eq.~\ref{Eq:sigreal} corrects for the wrong normalization applied to obtain $\sigma_{app}(E_\gamma)$. The second term represents the admixtures to the measured apparent cross section due to higher order radiation. Obviously there are unknown cross sections $\sigma(nE_\gamma)$ on the right side. Since their contributions are weighted by few-percent fractions $f_n$ of higher order radiation an iterative approach can be used to solve Eq.~\ref{Eq:sigreal}. It turns out that one iteration in which $\sigma(nE_\gamma)$ is replaced by $\sigma_{app}(nE_\gamma)$ is a sufficiently good approximation given the uncertainties of the energy dependent fraction $f_n(E_\gamma)$ of higher-order radiation.

For the present correction of W$^+$ cross sections only first-, second-, and third-order radiation are considered. Corrections are only necessary for the energy range of the first grating, i. e., in the energy range 16 to 80~eV.  The estimated total uncertainty of the resulting correction is assumed to be 50\% of the difference $\left|\sigma(E_\gamma) - \sigma_{app}(E_\gamma)\right|$ (see Sec.~\ref{sec:exp}).

%
%
%
%
\section*{References}

\providecommand{\newblock}{}
\begin{thebibliography}{}
\expandafter\ifx\csname url\endcsname\relax
  \def\url#1{{\tt #1}}\fi
\expandafter\ifx\csname urlprefix\endcsname\relax\def\urlprefix{URL }\fi
\providecommand{\eprint}[2][]{\url{#2}}

\end{thebibliography}


\begin{thebibliography}{10}
\expandafter\ifx\csname url\endcsname\relax
  \def\url#1{{\tt #1}}\fi
\expandafter\ifx\csname urlprefix\endcsname\relax\def\urlprefix{URL }\fi
\providecommand{\eprint}[2][]{\url{#2}}

\bibitem{Neu2013}
Neu R, Arnoux G, Beurskens M, Bobkov V, Brezinsek S, Bucalossi J, Calabro G,
  Challis C, Coenen J~W, {de la L}una E, {de V}ries P~C, Dux R, Frassinetti L,
  Giroud C, Groth M, Hobirk J, Joffrin E, Lang P, Lehnen M, Lerche E, Loarer T,
  Lomas P, Maddison G, Maggi C, Matthews G, Marsen S, Mayoral M~L, Meigs A,
  Mertens P, Nunes I, Philipps V, P\"{u}tterich T, Rimini F, Sertoli M, Sieglin
  B, Sips A~C~C, {van E}ester D, van Rooij G and {JET-EFDA Contributors} 2013
  {\em Phys. Plasmas\/} {\bf 20} 056111

\bibitem{Neu2003}
Neu R, Dux R, Geier A, Gruber O, Kallenbach A, Krieger K, Maier H, Pugno R,
  Rohde V, Schweizer S and {ASDEX Upgrade Team} 2003 {\em Fusion Eng. Des.\/}
  {\bf 65} 367

\bibitem{Mueller2015b}
M\"{u}ller A 2015 {\em Atoms\/} {\bf 3} 120

\bibitem{Rausch2011a}
Rausch J, Becker A, Spruck K, Hellhund J, {Borovik Jr} A, Huber K, Schippers S
  and M\"uller A 2011 {\em J. Phys. B: At. Mol. Opt. Phys\/} {\bf 44} 165202

\bibitem{Stenke1995b}
Stenke M, Hathiramani D, Hofmann G, Shevelko V~P, Steidl M, V\"{o}lpel R and
  Salzborn E 1995 {\em Nucl. Instrum. Methods B\/} {\bf 98} 138

\bibitem{Stenke1995c}
Stenke M, Aichele K, Harthiramani D, Hofmann G, Steidl M, V\"{o}lpel R and
  Salzborn E 1995 {\em J. Phys. B: At. Mol. Opt. Phys\/} {\bf 28} 2711--2721
  ISSN 0953

\bibitem{Schippers2011b}
Schippers S, Bernhardt D, M\"{u}ller A, Krantz C, Grieser M, Repnow R, Wolf A,
  Lestinsky M, Hahn M, Novotn\'{y} O and Savin D~W 2011 {\em Phys. Rev. A\/}
  {\bf 83} 012711

\bibitem{Krantz2014}
Krantz C, Spruck K, Badnell N~R, Becker A, Bernhardt D, Grieser M, Hahn M,
  Novotn\'{y} O, Repnow R, Savin D~W, Wolf A, M\"{u}ller A and Schippers S 2014
  {\em J. Phys. Conf. Ser.\/} {\bf 488} 012051

\bibitem{Spruck2014}
Spruck K, Badnell N~R, Krantz C, Novotn\'{y} O, Becker A, Bernhardt D, Grieser
  M, Hahn M, Repnow R, Savin D~W, Wolf A, M\"{u}ller A and Schippers S 2014
  {\em Phys. Rev. A\/} {\bf 90} 032715

\bibitem{Costello1991}
Costello J~T, Kennedy E~T, Sonntag B~F and Cromer C~L 1991 {\em J. Phys. B: At.
  Mol. Opt. Phys\/} {\bf 24} 5063

\bibitem{Sladeczek1995}
Sladeczek P, Feist H, Feldt M, Martins M and Zimmermann P 1995 {\em Phys. Rev.
  Lett.\/} {\bf 75} 1483

\bibitem{Trzhaskovskaya2010}
Trzhaskovskaya M, Nikulin V and Clark R~E~H 2010 {\em At. Data Nucl. Data
  Tables\/} {\bf 96} 1

\bibitem{Boyle1993}
Boyle J, Altun Z and Kelly H~P 1993 {\em Phys. Rev. A\/} {\bf 47} 4811

\bibitem{Ballance2015a}
Ballance C and McLaughlin B~M 2015 {\em J. Phys. B: At. Mol. Opt. Phys.\/} {\bf
  48} 085201

\bibitem{Mueller2011a}
M\"{u}ller A, Schippers S, Kilcoyne A~L~D and Esteves D 2011 {\em Phys. Scr.\/}
  {\bf T144} 014052

\bibitem{Mueller2012}
M\"{u}ller A, Schippers S, Kilcoyne A~L~D, Aguilar A, Esteves D and Phaneuf R~A
  2012 {\em J. Phys. Conf. Ser.\/} {\bf 388} 022037

\bibitem{Mueller2014c}
M\"{u}ller A, Schippers S, Hellhund J, Kilcoyne A~L~D, Phaneuf R~A, Ballance
  C~P and McLaughlin B~M 2014 {\em J. Phys. Conf. Ser.\/} {\bf 488} 022032

\bibitem{NIST2014}
{A E Kramida, Yu Ralchenko, J Reader, and NIST ASD Team (2014),} {NIST Atomic
  Spectra Database (version 5.2),} National Institute of Standards and
  Technology, Gaithersburg, MD, USA \urlprefix\url{http://physics.nist.gov/asd}

\bibitem{Covington2002a}
Covington A~M, Aguilar A, Covington I~R, Gharaibeh M~F, Hinojosa G, Shirley
  C~A, Phaneuf R~A, {\'A}lvarez I, Cisneros C, Dominguez-Lopez I, Sant'Anna
  M~M, Schlachter A~S, McLaughlin B~M and Dalgarno A 2002 {\em Phys. Rev. A\/}
  {\bf 66} 062710

\bibitem{Mueller2014b}
M\"{u}ller A, Schippers S, {Esteves-Macaluso} D, Habibi M, Aguilar A, Kilcoyne
  A~L~D, Phaneuf R~A, Ballance C~P and McLaughlin B~M 2014 {\em J. Phys. B: At.
  Mol. Opt. Phys.\/} {\bf 47} 215202

\bibitem{Fricke1980a}
Fricke J, M{\"u}ller A and Salzborn E 1980 {\em Nucl. Instrum. Methods\/} {\bf
  175} 379--384

\bibitem{Rinn1982}
Rinn K, M{\"u}ller A, Eichenauer H and Salzborn E 1982 {\em Rev. Sci.
  Instrum.\/} {\bf 53} 829

\bibitem{Phaneuf1999}
Phaneuf R~A, Havener C~C, Dunn G~H and M{\"u}ller A 1999 {\em Rep. Prog.
  Phys.\/} {\bf 62} 1143

\bibitem{Lu2006a}
Lu M, Gharaibeh M~F, Alna'washi G, Phaneuf R~A, Kilcoyne A~L~D, Levenson E,
  Schlachter A~S, M{\"u}ller A, Schippers S, Jacobi J, Scully S~W~J and
  Cisneros C 2006 {\em Phys. Rev. A\/} {\bf 74} 012703

\bibitem{Esteves2011}
Esteves D~A, Bilodeau R~C, Sterling N~C, Phaneuf R~A, Kilcoyne A~L~D, Red E~C
  and Aguilar A 2011 {\em Phys. Rev. A\/} {\bf 84} 013406

\bibitem{Ballance2006}
Ballance C~P and Griffin D~C 2006 {\em J. Phys. B: At. Mol. Opt. Phys.\/} {\bf
  39} 3617

\bibitem{Norrington1987}
Norrington P~H and Grant I~P 1987 {\em J. Phys. B: At. Mol. Phys.\/} {\bf 20}
  4869

\bibitem{Wijesundera1991}
Wijesundera W~P, Parpia F~A, Grant I~P and Norrington P~H 1991 {\em J. Phys. B:
  At. Mol. Opt. Phys.\/} {\bf 24} 1803

\bibitem{darc}
{DARC codes} \urlprefix\url{http://connorb.freeshell.org}

\bibitem{Fivet2012}
Fivet V, Bautista M~A and Ballance C~P 2012 {\em J. Phys. B: At. Mol. Opt.
  Phys.\/} {\bf 45} 035201

\bibitem{McLaughlin2012a}
McLaughlin B~M and Ballance C~P 2012 {\em J. Phys. B: At. Mol. Opt. Phys.\/}
  {\bf 45} 085701

\bibitem{McLaughlin2012b}
McLaughlin B~M and Ballance C~P 2012 {\em J. Phys. B: At. Mol. Opt. Phys.\/}
  {\bf 45} 095202

\bibitem{McLaughlin2014a}
McLaughlin B~M and Ballance C~P 2014 {\em { Petascale computations for
  large-scale atomic and molecular collisions Sustained Simulated
  Performance}\/} (New York: Springer) chap~15

\bibitem{McLaughlin2014b}
McLaughlin B~M, Ballance C~P, Pindzola M~S and M\"{u}ller A 2014 {\em {PAMOP:
  petascale atomic, molecular and optical collisions High Performance Computing
  in Science and Engineering}\/} (New York: Springer) chap~4

\bibitem{Hinojosa2012}
Hinojosa G, Covington A~M, Alna'Washi G~A, Lu M, Phaneuf R~A, Sant'Anna M~M,
  Cisneros C, \'{A}lvarez I, Aguilar A, Kilcoyne A~L~D, Schlachter A~S,
  Ballance C~P and McLaughlin B~M 2012 {\em Phys. Rev. A\/} {\bf 86} 063402

\bibitem{Dyall1989}
Dyall K~G, Grant I~P, Johnson C~T, Parpia F~A and Plummer E~P 1989 {\em Comput.
  Phys. Commun.\/} {\bf 55} 425

\bibitem{Parpia2006}
Parpia F~A, {Froese-Fischer} C and Grant I~P 2006 {\em Comput. Phys. Commun.\/}
  {\bf 175} 745

\bibitem{Grant2007}
Grant I~P 2007 {\em Relativistic Quantum Theory of Atoms and Molecules: Theory
  and Computation\/} Springer Series on Atomic, Optical, and Plasma Physics
  (New York: Springer)
  
\bibitem{Borovik2009a}
{Borovik Jr} A, M{\"u}ller A, Schippers S, Bray I and Fursa D~V 2009 {\em J.
  Phys. B: At. Mol. Opt. Phys\/} {\bf 42} 025203

\bibitem{Saghiri1999}
Saghiri A~A, Linkemann J, Schmitt M, Schwalm D, Wolf A, Bartsch T, Hoffknecht
  A, M{\"u}ller A, Graham W~G, Price A~D, Badnell N~R, Gorczyca T~W and Tanis
  J~A 1999 {\em Phys. Rev. A\/} {\bf 60} R3350

\bibitem{Renwick2009}
Renwick A~C, Bray I, Fursa D~V, Jacobi J, Knopp H, Schippers S and M\"{u}ller A
  2009 {\em J. Phys. B: At. Mol. Opt. Phys\/} {\bf 42} 175203

\bibitem{Mueller2014g}
M\"{u}ller A, {Borovik Jr} A, Huber K, Schippers S, Fursa D~V and Bray I 2014
  {\em Phys. Rev. A\/} {\bf 90} 010701(R)

\bibitem{Flambaum2015}
Flambaum V~V, Kozlov M~G and Gribakin G~F 2015 {\em Phys. Rev. A\/} {\bf 91}
  052704

\bibitem{Attwood1999}
Attwood D 1999 {\em {Soft X-rays and Extreme Ultraviolet Radiation - Principles
  and Applications}\/} (Cambridge University Press)

\bibitem{Emmons2005a}
Emmons E~D, Aguilar A, Gharaibeh M~F, Scully S~W~J, Phaneuf R~A, Kilcoyne
  A~L~D, Schlachter A~S, Alvarez I, Cisneros C and Hinojosa G 2005 {\em Phys.
  Rev. A\/} {\bf 71} 042704

\bibitem{Andersen2001b}
Andersen P, Andersen T, Folkmann F, Ivanov V~K, Kjeldsen H and West J~B 2001
  {\em J. Phys. B: At. Mol. Opt. Phys\/} {\bf 34} 2009

\bibitem{Mosnier2003a}
Mosnier J~P, Sayyad M~H, Kennedy E~T, Bizau J~M, Cubaynes D, Wuilleumier F~J,
  Champeaux J~P, Blancard C, Varma R~H, Banerjee T, Deshmukh P~C and Manson S~T
  2003 {\em Phys. Rev. A\/} {\bf 68} 052712

\end{thebibliography}

\providecommand{\newblock}{}

\end{document}